# Neutron investigation of the magnetic scattering in an iron-based ferromagnetic superconductor

Jeffrey W. Lynn,[1,*] Xiuquan Zhou,[2] Christopher K. H. Borg,[2] Shanta R. Saha,[3] Johnpierre Paglione,[3] and Efrain E. Rodriguez[2]

[1]NIST Center for Neutron Research, National Institute of Standards and Technology, Gaithersburg, MD 20899-6102

[2]Department of Chemistry and Biochemistry, University of Maryland, College Park, MD 20742

[3]Department of Physics, University of Maryland, College Park, MD 20742

Abstract

Neutron diffraction and small angle scattering experiments have been carried out on the double-isotopic polycrystalline sample ($^7$Li$_{0.82}$Fe$_{0.18}$OD)FeSe.  Profile refinements of the diffraction data establish the composition and reveal an essentially single phase material with lattice parameters of $a$= 3.7827 Å and $c$= 9.1277 Å at 4 K, in the ferromagnetic-superconductor regime, with a bulk superconducting transition of $T_C$ = 18 K.  Small angle neutron scattering (SANS) measurements in zero applied field reveal the onset of ferromagnetic order below $T_F \approx 12.5$ K, with a wave vector and temperature dependence consistent with an inhomogeneous ferromagnet of spontaneous vortices or domains in a mixed state.  No oscillatory long range ordered magnetic state is observed.  Field dependent measurements establish a separate component of magnetic scattering from the vortex lattice, which occurs at the expected wave vector.  The temperature dependence of the vortex scattering does not indicate any contribution from the ferromagnetism, consistent with diffraction data that indicate that the ordered ferromagnetic moment is quite small.

*corresponding author.  Jeffrey.Lynn@nist.gov

The magnetic properties of superconductors have a rich and interesting history.  Early work showed that even tiny concentrations of magnetic impurities destroyed the superconducting pairing through the exchange-driven spin depairing mechanism, prohibiting any possibility of cooperative magnetic behavior.[1]  The first exception to this rule was provided by the cubic rare-earth substituted CeRu$_2$ alloys [2-4], while the ternary Chevrel-phase (and related) superconductors (e.g. $R$Mo$_6$S$_8$, $R$=rare earth) provided the first demonstrations of long range magnetic order coexisting with superconductivity.[5,6]  The magnetic ordering temperatures were all quite low (~1 K), where electromagnetic (dipolar + London penetration depth) interactions play a dominate role in the energetics of the magnetic system. The vast majority of these materials order antiferromagnetically where coexistence of long range order with superconductivity was common, but these materials also provided the first examples of the rare occurrence of ferromagnetism and consequent competition with superconductivity in ErRh$_4$B$_4$,[7-10] HoMo$_6$S$_8$,[11-13] and HoMo$_6$Se$_8$.[14]  Antiferromagnetic order is found for all the rare earths in the cuprates, which exhibit similar low ordering temperatures.[15] In the borocarbide superconductors again of all the magnetic order is antiferromagnetic,[16] with the singular exception of $ErNi_2B_2C$ at low temperature [17, 18] where a net magnetization developed that resulted in the spontaneous formation of flux quanta (vortices).[19, 20]

For the high-$T_C$ superconductors of direct interest here, there have been no ferromagnets in either the cuprate or iron-based systems,[15, 21-23] with the possible exception of $RuSr_2GdCu_2O_8$ where canting of the Ru moments may produce a small net moment.[24-26] This situation changed recently with the reports by Nandi, *et al.* for doped $EuFe_2As_2$ [27] and Packmayr, *et al.* for $(Li_{1-x}Fe_xOH)FeSe$.[28] For this latter system $T_C$ can be as high as 43 K, together with the development of magnetic order below ~10 K with a suggested formation of a spontaneous vortex lattice. Here we report neutron diffraction and small angle neutron scattering (SANS) measurements on $(^7Li_{0.82}Fe_{0.18}OD)FeSe$, where the $^7Li$ isotope has been employed to avoid the neutron absorption of $^6Li$, and H has been replaced by D to avoid the huge nuclear incoherent cross section. We observe two separate components of magnetic scattering, one in zero applied field due to an inhomogeneous mixed state originating from either the spontaneous formation of vortices or ferromagnetic domains. With an applied field we observe the scattering from a well-developed vortex lattice.

The preparation of the powder sample was modified from a hydrothermal route reported in the literature.[28-32] The isotopically pure $^7LiOD$ precursor used for the synthesis was prepared by stoichiometric mixing of $^7LiCO_3$ (Sigma Aldrich, 99% for $^7Li$) and CaO (calcined from $CaCO_3$, Sigma Aldrich, 99%) in $D_2O$. The $CaCO_3$ precipitate was filtered, and $^7LiOD$ was crystallized by evaporation of the solution. The isotopically pure superconducting $^7Li_xFe_{(1-x)}ODFeSe$ ($x \approx 0.18$) sample was prepared by hydrothermal reaction of 0.5 g Fe powder (Alfa Aesar, 99.9%), 1.5 g selenourea (Sigma Aldrich, 98%) and 3.5 g $^7LiOD$ in 9 mL of distilled $D_2O$ (Oxford Isotope, 99.9%) in a stainless steel autoclave at 150 °C for 3 days. The autoclave was opened in an argon-filled glove bag, and the shiny black precipitate was washed with $D_2O$ and centrifuged several times until the supernatant was clear. The remaining product was collected, vacuum dried, and stored in a nitrogen-filled glove box. Magnetic susceptibility measurements were carried out using a magnetic property measurement system (Quantum Design MPMS). Both field-cooled (FC) and zero-field-cooled (ZFC) magnetic susceptibility measurements were taken from 2-320 K in DC mode with an applied magnetic field of 1 mT. Room temperature powder X-ray diffraction (PXRD) data were collected on a Bruker D8 X-ray diffractometer with Cu Kα radiation, λ = 1.5418 Å. All the neutron work was carried out at the NIST Center for Neutron Research. Low temperature diffraction data were collected on the BT-1 high resolution powder neutron diffractometer (PND) with the Cu(311) monochromator ( λ = 1.540 Å). Scans were taken on a 1 g sample at 4 K and 150 K. Both LeBail fits to the powder XRD and the Rietveld refinements with PND were carried out with the TOPAS 4.2 software.[33] High intensity-coarse resolution diffraction measurements were carried out on the BT-7 spectrometer using the position sensitive detector to search for magnetic Bragg peaks.[34] SANS measurements were carried out on the NGB30SANS (no magnetic field) and NG7SANS (field-dependent work) spectrometers.

Figure 1 shows the neutron diffraction pattern for the $^7Li_xFe_{(1-x)}ODFeSe$ sample collected at 4 K, which is the expected crystal structure analogous to that prepared in $H_2O$.[28,30,32,35] Profile refinement of the tetragonal superconducting phase gave lattice constants a = 3.7827(1) Å, c = 9.1277(3) Å, and with x = 0.183(6). A table of the refined parameters are provided in the SI. The top and bottom ticks indicate the superconducting phase and about 10% $Li_2CO_3$ impurity phase, respectively, and the * indicates an FeSe impurity estimated at less than 1 wt. %, whose peaks are considerably broader than instrumental resolution. Neither impurity affects the superconducting properties.[30] The crystal structure is shown in the inset, where the FeSe layer is isostructural to bulk FeSe and is presumed to be where the superconducting pairing occurs. We find that both sites are fully occupied. In the $(Li_{1-x}Fe_x)OD$ layer, where the ferromagnetic order is expected to occur, we have x=0.183(6) for this sample.[36] The overall crystal structure is in excellent agreement with previous reports, although we find that the replacement of H by D does modify to some degree the lattice parameters and composition dependence of the magnetic-superconducting phase diagram. For this value of x the measured superconducting transition $T_c$ = 18 K as shown in Fig. 2, which is in the range of $T_c$'s reported in the study by Sun, *et al.* (15 to 40 K).[32] At lower temperatures we observe hysteresis develop in the field-dependent data consistent with the development of a net magnetization in the system.

High intensity powder diffraction measurements were carried out on BT-7 to search for magnetic Bragg peaks. In the case of antiferromagnetic ordering where the magnetic peaks are distinct from the structural peaks, an ordered moment somewhat below 0.1 $\mu_B$ can be detected without too much difficulty, such as was done for a 1 g LaFeAsO sample with an ordered moment of 0.3 $\mu_B$.[37] For the present system no evidence for antiferromagnetic ordering was found.[35] For a ferromagnet the magnetic Bragg peaks always occur at the same Bragg positions as the structural peaks, and we were also unable to detect any ordered ferromagnetic moment (see the SI). This perhaps is not surprising given that the ferromagnetism is expected to arise in the Li-Fe layer; with an expected moment of less than 1 $\mu_B$/Fe and only 18% of the sites occupied, the site-averaged ferromagnetic moment will be difficult to observe in powder diffraction.

An alternative method for detecting long wavelength oscillatory magnetic order (such as in $HoMo_6Se_8$[14]) or pure ferromagnetism (such as below the reentrant transition in $HoMo_6S_8$[12,13]) is via small angle neutron scattering. To explore the small angle magnetic scattering, data were taken with three different instrumental configurations using a closed cycle refrigerator (no field), with the temperature varying from 2.5 K to 37.5 K, all with no guides in the incident beam. Initial data employed a wavelength $\lambda$=4.5 Å and a detector position of 7 m, which spanned the wave vector Q range from >~0.01 Å$^{-1}$ (where the beam stop that absorbs the undeviated neutrons blocks the signal) to 0.11 Å$^{-1}$. In this wave vector regime no significant change in scattering was observed from 2.5 K to 37.5 K. To explore smaller Q's we increased the wavelength to $\lambda$ = 9 Å and increased the detector position to 13.5 m, which permitted the scattering to be measured in the wave vector range of ~2.5 10$^{-3}$ Å$^{-1}$ to 2.5 10$^{-2}$ Å$^{-1}$. No significant change in the scattering was observed between 25 K and 37 K, and thus these data were averaged and used as background. Fig. 3a shows the wave vector dependence of the difference scattering in this regime at several temperatures of interest, with data being taken between 2.5 K and 37.5 K in steps of 2.5 K. Between 12.5 K and 25 K again very little magnetic scattering is observed, while for lower temperatures we observe a rapid increase in intensity with decreasing T. At each temperature this scattering increases monotonically with decreasing Q, but otherwise does not appear to change its shape. In particular, it does not exhibit a peak as would be expected for a long wavelength oscillatory magnetic state such as observed in $ErRh_4B_4$,[9,10] $HoMo_6S_8$,[13] and $HoMo_6Se_8$.[14] Additional data were taken with a wavelength of $\lambda$=15 Å at 2.5 K and 25 K to push the measurement cutoff to 0.0015 Å$^{-1}$, but the shape of the scattering was unchanged. Fig. 3b shows the integrated intensity for this scattering as a function of temperature. We see that there is a well-defined onset of scattering at the onset of ferromagnetism at ~12 K . This scattering could originate from the spontaneous formation of vortices. With a randomly oriented powder and no field applied, any vortices that form will be in a mixed state. Whether they organize into a well-defined vortex lattice or are just individual vortices, they will be oriented randomly in the powder and the overwhelming majority won't be oriented to (coherently) Bragg diffract. Nevertheless there will be magnetic scattering from individual vortices, and this could explain the SANS scattering shown in Fig. 3. An alternative possibility is to note that the Q and T dependence of this scattering is quite similar to the expected scattering from ferromagnetic domains/domain walls, such as has been observed in the polycrystalline $Tl_2Mn_2O_7$ ferromagnet.[38]

To investigate the field dependence of this small angle magnetic scattering as well as search for scattering from the field-induced vortex lattice, we carried out additional SANS measurements under identical instrumental conditions in a horizontal field (parallel to the incident neutrons) superconducting SANS magnet. For the field-dependent measurements the sample was pressed into a pellet to prevent the grains from possibly changing orientation with field. Fig. 4a shows the net intensity at 5 K upon cooling in an applied field of 0.4 T. We see that in comparing with the data in Fig. 3a, the ferromagnetic scattering has shifted to smaller Q, as might be expected since in this layered superconductor the spontaneous vortices only in some appropriately-aligned grains will align with the field, or in the case of ferromagnetic domains they would grow larger, with fewer of them. We also see a well-defined peak at larger Q. A least-squares fit to a (resolution-limited) Gaussian peak yields a position Q = 0.0078(3) Å$^{-1}$. The expected position for a triangular vortex lattice is given by

$$Q_{10} = 2\pi \sqrt{\frac{2B}{(\sqrt{3})\phi_0}}$$

where $\phi_0 = 2.068 \times 10^5$ T Å$^2$ is the flux quantum and B is the internal field.[39] For an applied field of 0.4 T the calculated position is Q= 0.0077 Å$^{-1}$ (assuming a spherical shape), in excellent agreement with the measurement.

The temperature dependence of the vortex scattering was determined by integrating the net scattering over the Q range where the peak occurs, and is shown in Fig. 4b. We develop a signal below $T_C$ as expected. There should also be a contribution from the field-induced moment above the ferromagnetic transition. In the ferromagnetic state we expect an additional contribution from the internally generated magnetic flux. Any internally generated magnetization would cause both a shift of the vortex peak to larger Q, and an increase in intensity due to the increase in the number of vortices, as found for $ErNi_2B_2C$.[19,20] Neither trend is observed in these data, again indicating that the ferromagnetic moment is quite small.

For ferromagnetic superconductors like $ErRh_4B_4$, $HoMo_6S_8$ and $HoMo_6Se_8$ the magnetization that develops in the superconducting state competes with the Meissner screening through the London penetration depth. For the first two materials, initially this competition results in a long wavelength (incommensurate) oscillatory magnetic order that coexists with superconductivity, but as the magnetization increases the superconducting state is quickly quenched and a strongly first-order transition occurs to Q=0, i.e. pure ferromagnetism. For $HoMo_6Se_8$ where the ordered moment is smaller than $HoMo_6S_8$ and $T_C$ is higher, the wavelength of the oscillatory state increases as the magnetization increases, moving towards Q=0, but superconductivity persists and the long range ordered magnetic ground state remains oscillatory. The present system certainly behaves differently in that we do not see any oscillatory magnetic order, thus leaving the other possibility that vortices spontaneously form, as has been observed for ErNi$_2$B$_2$C and has been anticipated to be the case for the *5f* ferromagnetic superconductors UGe$_2$,[40] and URhGe ,[41] UCoGe [42] as well as for the present system [28] and doped EuFe$_2$As$_2$.[27]  In the U materials the ordered moment is quite small (a few hundredths µ$_B$), apparently comparable to the present system.  If a peak at finite Q in the zero-field scattering had been observed for the present system, these two cases would have been easily distinguished, since with increasing ferromagnetic moment the oscillatory peak would have moved to smaller Q like HoMo$_6$Se$_8$,[14] while the shift would have been to larger Q for the spontaneous vortex case. If we assume this low-Q scattering originates from a spontaneous vortex lattice that occurs at smaller wave vectors than accessible in the present measurements, this places an upper limit of ~0.01 T to the magnetic field generated by the ferromagnetism, which would correspond to a site-averaged ordered moment below ~0.09 µ$_B$.  For the field-induced case, on the other hand, the 'spontaneous vortex scattering' will occur together with the field-induced vortex peak in Fig. 4a, and there should be no small-Q magnetic component unless the magnetic anisotropy is too large or the vortices are strongly pinned.  This leaves ErNi$_2$B$_2$C as the only material where vortices have been observed to spontaneously form due to an internally generated magnetic field, albeit with a field applied.[19,20]  Consequently a true spontaneous vortex lattice—formed in the absence of an applied field—still remains to be observed in any ferromagnetic superconductor.[27, 40-46]  What is clear is that Li$_{1-x}$Fe$_x$OHFeSe is a fascinating ferromagnetic superconductor, and further measurements to investigate the magnetic order and vortex formation in greater detail when appropriate single crystal samples become available should prove very interesting.

Acknowledgments

We thank Cedric Gagnon for his assistance with the SANS measurements, and Qiang Ye for his assistance with the 9 T superconducting magnet system.  Research at the University of Maryland was supported by NSF Career DMR-1455118.  The NGB30SANS spectrometer is supported in part by NSF, DMR-0944772.  The identification of any commercial product does not imply endorsement or recommendation by the National Institute of Standards and Technology.

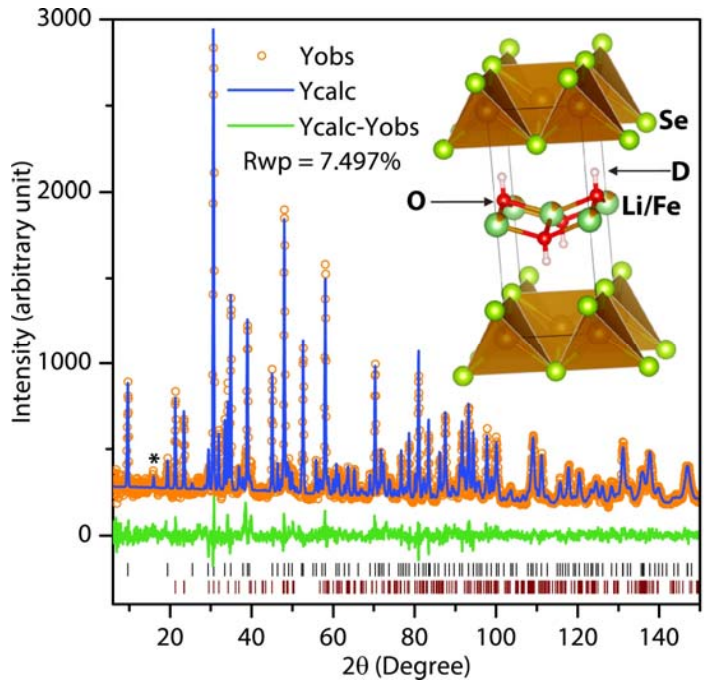

Figure 1. (color online) Rietveld refinement of the neutron diffraction pattern for the $^7$Li$_x$Fe$_{(1-x)}$ODFeSe ($x \approx 0.183(6)$) sample collected at 4 K using the Cu(311) monochromator ($\lambda = 1.540$ Å) at the BT-1 diffractometer. Structural refinement of the tetragonal superconducting phase gave lattice constants $a = 3.7827(1)$ Å and $c = 9.1277(3)$ Å. Top and bottom ticks indicate the superconducting phase and about 10% Li$_2$CO$_3$ impurity phase, respectively. The * indicates a (Fe-Se) impurity estimated at less than 1 wt. %, whose peaks are considerably broader than instrumental resolution. The tetragonal unit cell is shown as an inset.

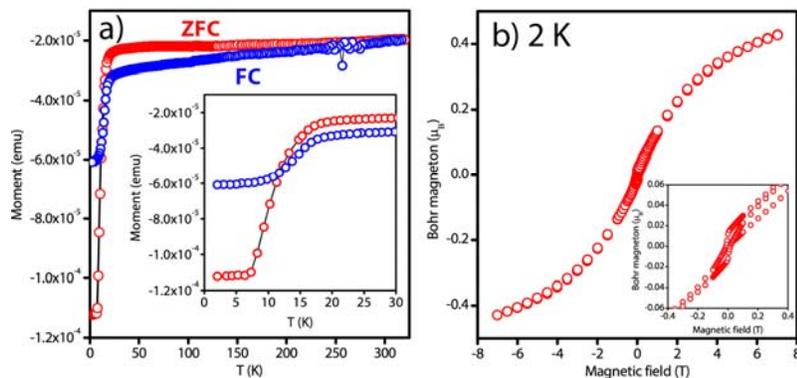

Figure 2. (color online) a) Temperature dependence of magnetic moment of the $^7$Li$_x$Fe$_{(1-x)}$ODFeSe ($x \approx 0.18$) sample measured in a SQUID magnetometer in an applied field of 1 mT. The zero field cooled data (ZFC) were taken upon heating and the field cooled (FC) data upon cooling. Inset shows the data below 30 K revealing the $T_c \approx 18$ K. Note that 1 emu = $10^{-3}$ A m$^2$. b) Magnetic field dependence of isothermal magnetization per formula unit of the same sample measured in hysteresis mode and ZFC. The inset expands the low field region to highlight the hysteresis, as expected for a ferromagnet.

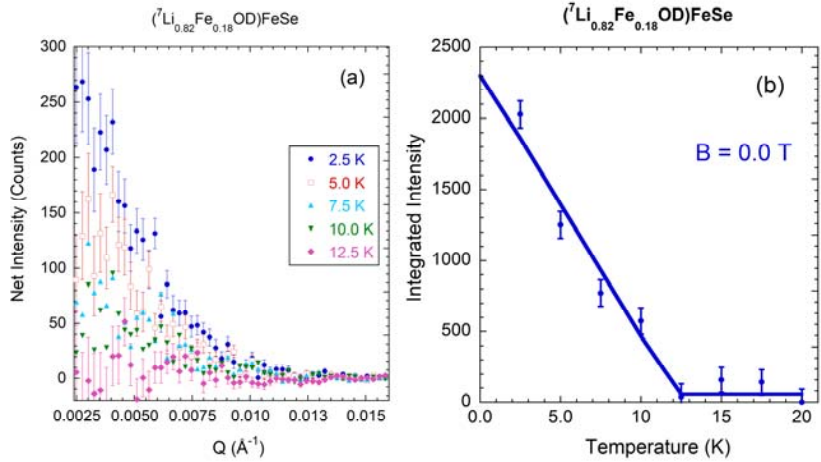

Figure 3. (color online) (a) Magnetic scattering as a function of wave vector Q for several temperatures. No magnetic field is applied. At 12.5 K very little magnetic scattering is observed, as is the case at higher temperatures (see text). Below the ferromagnetic transition magnetic scattering intensity develops, which increases monotonically with both decreasing Q and T. No peak in this Q range is observed, ruling out the formation of a long range ordered oscillatory magnetic state. (b) Integrated intensity as a function of temperature, revealing a magnetic transition temperature of ~12.5 K. Uncertainties are statistical in origin and represent one standard deviation.

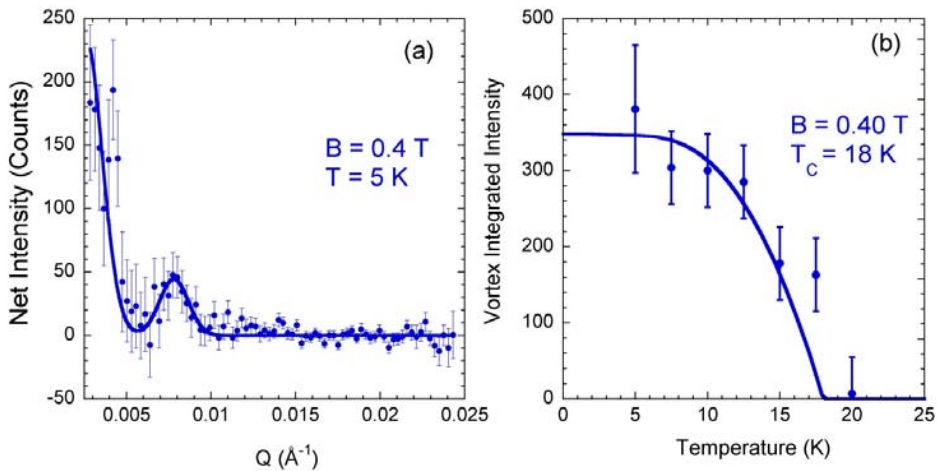

Figure 4. (color online) (a) Magnetic intensity at 5 K after cooling from 25 K in an applied field of 0.4 T. The ferromagnetic scattering has shifted to smaller Q, indicating that the length scale has increased or the strength has decreased, as would be expected when a field is applied. The peak at Q = 0.0077 Å$^{-1}$ is due to vortex scattering. (b) Integrated intensity of the vortex scattering as a function of temperature. The onset of scattering occurs at $T_C$ = 18 K. No evidence of the ferromagnetic ordering is observed, indicating that the ordered moment is small (see text).